# HIGH ENERGY NEUTRINOS FROM QUASARS

F.W. STECKER and M.H. SALAMON*
*Laboratory for High Energy Astrophysics*
*NASA/Goddard Space Flight Center*
*Greenbelt, Maryland 20771*

**Abstract.** We review and clarify the assumptions of our basic model for neutrino production in the cores of quasars, as well as those modifications to the model made subsequently by other workers. We also present a revised estimate of the neutrino background flux and spectrum obtained using more recent empirical studies of quasars and their evolution. We compare our results with other theoretical calculations and experimental upper limits on the AGN neutrino background flux. We also estimate possible neutrino fluxes from the jets of blazars detected recently by the EGRET experiment on the Compton Gamma Ray Observatory. We discuss the theoretical implications of these estimates.

## 1. Introduction

Quasars are the most powerful emitters of radiation in the known universe. These remarkable objects are presumably fueled by the gravitational energy of matter falling onto a supermassive black hole at center of the quasar core, though the mechanism responsible for the efficient conversion of gravitational to observed luminous energy is not yet known. If this conversion occurs through the acceleration of high energy cosmic rays, their interactions can lead to the copious production of hadronic and subsequent leptonic high energy byproducts such as neutrinos. It is in this context that quasars have long been considered as potential sources of high energy cosmic ray neutrinos (Berezinsky 1977; Eichler 1979; Silberberg and Shapiro 1979; Berezinsky and Ginzburg 1981).

In related work, Biermann and Strittmatter (1987) suggested that protons could be accelerated to ultrahigh energies in the hot spots found in the jets of radio galaxies (see also Rachen, Stanev and Biermann 1993). Subsequent X-ray production at these sites was considered by Mannheim, Kruells and Biermann (1991). High energy $\gamma$-ray and neutrino production in jets was explored by Mannheim, Stanev and Biermann (1992), following the important discovery of strong high energy $\gamma$-ray emission from the blazar 3C279 (Hartman, *et al.* 1992). New results on the detection of large fluxes of high energy $\gamma$-rays from $\sim 40$ such sources, obtained using the EGRET detector aboard the Compton Gamma Ray Observatory (Fichtel, *et al.* 1994), have made the study of these high energy sources even more interesting.

An important step in modeling quasar neutrino production process was made when Begelman, Rudak and Sikora (1990) and Stecker, *et al.* (1991)

pointed out the importance of photomeson production off thermal UV accretion disk photons as the most significant production channel.

In our work (Stecker, *et al.* 1991), we pointed out that following photopion production, pair cascading off the intense quasar photon fields would both destroy accompanying high energy $\gamma$-rays and produce the quasar's X-ray emission (*e.g.*, Done and Fabian 1989). Thus, by normalizing to the X–ray emission, we argued that active galaxies could be by far the strongest sources of ultrahigh energy neutrinos in the universe, producing fluxes detectable on the Earth with present technology. Measurement of these neutrino fluxes and spectra could thus provide key information as to the fundamental astrophysical nature of quasars.

We further pointed out that by integrating over all quasars in the Universe, a large and measurable isotropic neutrino background would be produced. Because of the degradation of $\gamma$-rays to X–ray energies, the energy in quasar background neutrinos could be comparable to that of the cosmic X–ray background rather than the much smaller energy of the cosmic $\gamma$-ray background. We also noted that radio quiet quasars are a factor of 10 more abundant than their radio loud counterparts (Sramek and Weedman 1978; Kellerman, *et al.* 1989). They thus take on corresponding significance when considering their contribution to the high energy cosmic neutrino background.

Szabo and Protheroe (1994) have followed up on this basic model with independent calculations. They have examined a series of models and assumed that nucleons are trapped indefinitely in the quasar core. In such models, the nucleons eventually lose energy and interact with the X–rays in the core as well as the UV photons, even though the X–ray field of the core is optically thin to nucleons. The result is a much higher flux of lower energy neutrinos than we calculated. The difference between the predicted event rates in neutrino detectors between the two calculations is not as pronounced. This is because the detectability of high energy neutrinos ($\propto \sigma_{\nu N} \times$ range of produced muon) is $\propto E_\nu \ln(E_\nu)$ in this energy range.

An extensive discussion of quasar neutrino production took place at an historic workshop in Hawaii, with almost all of the above-mentioned workers taking part. The reader is referred to the proceedings of this workshop (Stenger, *et al.* 1992) for more background into this subject. In that workshop, Stecker, *et al.*, Szabo and Protheroe, Biermann, and Sikora and Begelman all gave estimates of the neutrino background from AGN.

In this paper, we will update our model using new information on quasar evolution and will also consider neutrino production in blazar jets. We will then compare our results with other work. We will also review the present experimental upper limits on cosmic high energy neutrino fluxes and their implications.

## 2. The AGN Phenomenon

The standard "AGN paradigm" assumes that infalling matter forms an accretion disk around a black hole. It has been suggested by Kazanas and Ellison (1986) that an accretion disk shock will form at a distance of several Schwartzschild radii ($R_s = 2GM/c^2$) from the black hole. First–order Fermi acceleration of charged particles at this shock then converts a significant fraction of the gravitational energy into highly relativistic particles which have the characteristic hard power law spectrum typical of shock acceleration. Other possible energy generation mechanisms involve electromagnetic acceleration by black hole accretion disk dynamo models. In any model where high energy cosmic rays are produced at the core of the quasar, such cosmic rays (even neutrons) will be trapped by the intense radiation fields. Thus, a necessary consequence of *any* acceleration mechanism in which protons reach high energies is the production of high energy neutrinos through the collisions of these particles with the intense photon fields in quasars.

Recent observational studies bearing on quasar evolution and theoretical models of quasar formation and evolution can be used to guide one in choosing the basic physical parameters needed to calculate neutrino fluxes from individual quasars and also to calculate the neutrino background from all "unresolved" quasars. There is empirical justification for assuming that all quasars run at some constant high efficiency. Quasar evolution appears to be characterised a phase of emission near the maximum allowed rate. This is quantified in terms of the the Eddington limit ($L_{\rm edd} = 4\pi GM m_p c/\sigma_T$ where $M$ is the mass of the black hole), the maximum steady state luminosity that can be produced before radiation pressure disrupts the accretion flow. Studies have shown that quasars and Seyfert galaxies are emitting at a substantial fraction of the Eddington luminosity at all observed redshifts. For quasars, this fraction, which we will call the "Eddington fraction", appears to be usually above $10^{-1}$, and for the less luminous Seyfert galaxies, above $10^{-2}$ with a substantial number above $10^{-1}$ (Padovanni and Rafanelli 1988).

These results are in accord with theoretical beliefs that as long as there is an adequate supply of gas available, the luminosity produced by the "central engine" of the quasar should approach $L_{\rm edd}$. This also implies that the bright phase of a quasar *should not be substantially longer than the Eddington time*, $t_{\rm edd} = Mc^2/L_{\rm edd} = 4.4 \times 10^8$ yr. Thus, it appears that continuous models of quasar evolution are ruled out (Cavaliere and Padovani 1988).

Haehnelt and Rees (1993) have pointed out that the similarity between the timescale $t_{\rm edd}$ and the evolutionary timescale for the hierarchial buildup of galaxy size masses at high redshifts allows for a natural explanation of the strong evolution of the quasar population. In their work, they point out that if the characteristicly large Eddington fraction of quasars is insensitive to redshift, the only way to understand the rapid increase of quasar luminosity with redshift is to allow that the black holes which are active at low redshifts are different sources with typically smaller masses than those which were emitting at higher redshifts.

We thus arrive at a picture where all black holes in their active phase had an adequate supply of gas to accrete near their Eddington limit so that the Eddington fraction is both large and roughly constant with redshift, as we assumed in our original paper (Stecker, *et al.* 1991). In this scenario, typically the larger mass black holes turned on and used up their supply of gas early, to be replaced in later generations by smaller mass black holes, which, in turn, used up their gas. Such a picture would naturally explain why the quasar X–ray spectral index is no different at $z = 3$ than at $z = 0$, even though their mean luminosity at $z = 3$ is ∼30 to 40 times larger than that at $z = 0$ (Elvis 1994). The short lifetime scenario allows for both density and luminosity evolution of the quasar population. However, the ROSAT results (Boyle, *et al.* 1993) appear to be consistent with the simple assumption of pure luminosity evolution, and we will use that analysis accordingly in updating our calculation of the neutrino background from quasars.

## 3. Cosmic Ray Interactions in Quasar Cores

Let us assume that cosmic rays are accelerated in the centers of active galaxies (see section 4) and examine their expected interactions and energy losses, including those which lead to $\gamma$-ray and neutrino production. First, we consider the photon environment.

A characteristic quasar spectrum has comparable amounts of power in each logarithmic energy interval, but the spectrum is not completely flat. Typically, the most prominent feature of the spectrum is the "blue bump", which has a pronounced peak in the hard UV and is thought to be thermal emission from the black hole accretion disk. This feature is thought to arise from the release of energy through viscous processes in the accretion disk (*e.g.*, Laor 1990).

There is also usually a thermal feature in the IR region, thought to be due to dust intercepting and reradiating the UV emission. This hypothesis is supported by the fact that the observable "blue bump" is not as pronounced in AGN which have prominent thermal IR features. Dust can only exist outside of the nucleus, as it is destroyed by sputtering, so the IR spectrum must arise at a large distance from the "central engine" (Barvainis 1987). The lack of variability in the IR spectrum also implies that the emission is from a large region (Edelson and Malkan 1987).

Between the IR and UV emission features there may also be a further continuum component (Loska and Czerny 1990; Laor 1990), but this component appears to be not associated with the very central core (Done *et al.* 1990).

The most rapidly variable part of the spectrum is the X–ray region. Its emission is similar in luminosity to that of UV bump but its rapid time variability indicates that the X–rays are produced in a smaller region than the UV flux (Lawrence, *et al.* 1987; McHardy and Czerny 1987; McHardy 1988). Thus, *in the very central regions, only the UV and X–ray photon fields are of importance.*

The lack of strong X-ray absorption features in quasar spectra (Mushotzky 1982; Turner and Pounds 1989) implies that the secondary X–rays are produced in regions of low column density. This puts strict limits on the amount of target gas for *pp* interactions, and one finds that the very large photon density in the quasar core makes $p\gamma$ interactions the dominant energy loss mechanism. This is a threshhold interaction, and the dominant $\nu$ production channel at energies just above the threshold is $\gamma p \to \Delta \to n\pi^+$ (Stecker 1968).

The optical depth to photomeson production for protons of energy $E_p$ can be approximated by $\tau(E_p) \sim \epsilon n(\epsilon) \sigma_o R$ where $\sigma_o$ at the $\Delta$ resonance peak is $\approx 5 \times 10^{-28}$ cm$^2$ and $n(\epsilon)$ is the differential number density of photons at the resonance energy in cm$^{-3}$ eV$^{-1}$. An estimate for $n(\epsilon)$ from the accretion disk can be found by assuming that all sources run at some constant efficiency (see Section 2). Fitting accretion disk models to the UV spectrum (Laor 1990) gives a typical luminosity of $(0.03 - 0.1)L_{\text{edd}}$ ergs s$^{-1}$. We have assumed a "blue bump" UV luminosity of $\sim 0.05 L_{\text{edd}}$ is emitted isotropically. Given the weak dependence of the characteristic accretion disk temperature on black hole mass and the very similar UV bump shapes seen in the observational data, we can assume a luminosity independent 'generic' UV quasar spectrum. The quasar X–ray spectrum is also fairly universal, typically being an $\epsilon^{-1.7}$ power law above 2 keV (Mushotzky 1982; Turner and Pounds 1989), with a cutoff in the MeV energy range (Rothschild, *et al.* 1983). The total X–ray luminosity is roughly the same as that in the UV bump *i.e.* $L_{\text{x}} \approx L_{\text{UV}} \approx 0.05 L_{\text{edd}}$. Normalizing so that $L \sim 4\pi R^2 c \int \epsilon n(\epsilon) d\epsilon = 0.1 L_{\text{edd}}$, we can calculate the mean UV and X–ray photon density within the UV emission region of radius R. We thus obtain

$$n(\epsilon) \approx \frac{5 \times 10^{14}}{L_{45}} \text{cm}^{-3}\text{eV}^{-1} \begin{cases} 0.165\epsilon & \epsilon < 1 \text{ eV} \\ 0.165\epsilon^{-0.9} & 1 < \epsilon < 40 \text{ eV} \\ 5.35 \times 10^{-5}\epsilon^2 e^{-\epsilon/15} & 40 < \epsilon < 192 \text{ eV} \\ 4.15 \times 10^{-2}\epsilon^{-1.7} & 192 < \epsilon < 10^6 \text{ eV} \end{cases} \quad (1)$$

where $L_{45}$ is the total UV luminosity in units of $10^{45}$ ergs s$^{-1}$. The $1/L$ scaling of $n(\epsilon)$ follows from the linear scaling of both $R$ and $L$ with $M$ (*i.e.* with $L_{\text{edd}}$ and $R_s$ above.)

The optical depth from the "blue bump" UV photons is $\gg 1$. The optical depth is independent of the luminosity since $\tau(E_p) \propto n(\epsilon) R$ and $n(\epsilon) \propto 1/L$ while $R \propto L$. The high optical depth of the UV photon field implies that the secondary neutrons also will not generally escape the luminous core region, so that comparable amounts of power are generated via $n\gamma$ collisions. With roughly half of the energy loss going into $\pi^0$'s and the other half into $\pi^{\pm}$'s, the luminosity for $\nu_\mu + \overline{\nu}_\mu$ is $\sim 0.4 L_\mathrm{x}$, and for $\nu_e + \overline{\nu}_e$ is $\sim 0.2 L_\mathrm{x}$. As the quasar photon spectrum drops exponentially between 40 and 200 eV, it is overwhelmingly the UV bump photons which interact with the high-energy nucleons. We note that for this typical quasar spectrum, the X–rays are not optically thick.

## 4. Proton Acceleration in Quasar Accretion Disk Shocks

It has been suggested by Kazanas and Ellison (1986) that protons can be accelerated to very high energies by diffusion across the strong shock generated at the inner edge of the black hole accretion disk. First-order Fermi acceleration of protons in strong (non-relativistic) shocks produces a power-law proton energy spectrum $\propto E_p^{-2}$ up to a maximum energy $E_\mathrm{max}$. This maximum proton energy is determined by equating the proton acceleration time, $t_\mathrm{acc}(E_p)$, with the $p\gamma$ lifetime, $t_{p\gamma}(E_p)$. The proton acceleration time is given by $t_\mathrm{acc}(E_p) \approx 2.2 \times 10^{-4}(R_\mathrm{shock}/R_s)(E_p/m_p)B^{-1}$ (Kazanas and Ellison (1986). The $p\gamma$ lifetime is given by $t_{p\gamma}(E_p) \approx (N_\gamma \sigma_{p\gamma} c\kappa)^{-1}$, where $\kappa$ is the mean elasticity and $N_\gamma$ is the number density of photons above the photoproduction threshhold. The shock radius, $R_\mathrm{shock}$, is fixed at $\sim 10 R_s$ by our assumption that the X–ray luminosity is $0.05 L_\mathrm{edd}$. The magnetic field $B$ is taken to be $10^3\, L_{45}^{-1/2}$ G, which assumes approximate equipartition of the magnetic energy density with the quasar accretion disk "UV bump" photon density. (Note that this is a change from our earlier work (Stecker, et al. 1991) where we just assumed a luminosity independent value of $B = 10^3$ G.) Setting the loss time equal to the acceleration time, and noting that $N_\gamma \propto 1/L$ (see section 3), we obtain $E_\mathrm{max}$ given by equation (2). For $L_{45} > 0.1$, $E_\mathrm{max} \propto L^{1/2}$, a result similar to that obtained by Szabo and Protheroe (1994) using this same assumption about the $B$ field.

$$E_\mathrm{max} = 10^9 \mathrm{GeV} \cdot \begin{cases} 0.25 L_{45}^{0.375}, & L_{45} \leq 1 \\ 0.25 L_{45}^{0.5}, & 1 < L_{45} \leq 10^2 \\ 25, & L_{45} > 10^4 \end{cases} \quad (2)$$

For all luminosities, the maximum proton energy is limited by $p\gamma$ interactions rather than by the scale size of the shock region.

## 5. The Production of Ultrahigh Energy Neutrinos

Given a power law (roughly $\propto E^{-2}$) proton spectrum, the $p\gamma$ pion producing interactions occur mostly close to the threshhold energy. The dominant $\nu$ production channel at energies just above the threshold is $\gamma p \rightarrow \Delta \rightarrow n\pi^+$ (Stecker 1968). The cross-section for this reaction peaks at photon energies of $\epsilon' = 0.35 m_p c^2$ in the proton rest frame. In the observers frame $\epsilon E_p = 0.35$ eV-EeV with $E_p$ in EeV and $\epsilon$ in eV. For UV bump photons, with a mean energy of ∼30 eV, this translates to a characteristic proton energy of $E_c \sim 10^7$ GeV.

It is possible to estimate the isotropic X–ray flux from the core of the closest and brightest quasar, 3C273. Its X–ray spectrum is observed to exhibit an iron line at 6.4 keV of equivalent width ∼ 50 eV (Turner, *et al.* 1990). The currently accepted explanation for this line in quasars is fluorescence from the X–ray illuminated accretion disk, so that much of the X–radiation is probably unbeamed. The isotropic X–ray emission in the 2–10 keV band is then ∼ $10^{46}$ ergs s$^{-1}$, giving a total X–ray flux of ∼ $10^{47}$ ergs s$^{-1}$.

It now appears that, in the case of the *Seyfert galaxy* NGC 4151, at least some of the X-ray flux and perhaps most or all of the X-ray flux may be of thermal origin with Comptonization playing a major role (Zdziarski, Lightman and Maciolek-Niedzwiecki 1993; Titarchuk 1994). Such a contribution to a *quasar* X-ray spectrum may correspondingly lower our estimate of its neutrino flux, since the neutrino flux in our model is proportional only to the non-thermal AGN component of the X–ray emission. (See, however, Zdziarski *et al.*, 1994; 1995.) To take account of this eventuality, we will assume that, on average, only 30% of quasar X-ray emission is non-thermal in normalizing our calculations. Using this estimate, we calculated the neutrino spectrum of 3C273. The result shown in Figure 1 was calculated using Monte Carlo techniques.

The mean pion energy is $<E_\pi> \sim 0.2 E_p$, while the mean neutrino energy is $\sim E_\pi/4 \sim 0.05 E_p$ (Stecker 1979). Thus the neutrino spectrum from proton interactions with the UV bump extends from $E_c/20$ to $E_{\max}/20$ *i.e.* from ∼ $10^6$ to ∼ $10^8$ GeV, with a spectral index of ∼ 2, mirroring that of the proton spectrum. Above ∼ $10^9$ GeV, the cutoff in the proton spectrum leads to a cutoff in the neutrino spectrum. Below $E_c/20$ relativistic kinematics give rise to a flat $\nu$ energy distribution (Stecker 1979).

It is probable that there is a range of X–ray compactness in quasars, and hence a range of optical depths to the $p\gamma$ process. For more compact quasars, the maximum energy of the protons would be lower, with the $p\gamma$ interactions with X–rays determining the cutoff (Sikora, *et al.* 1987). Interactions with X–rays will also result in the more copious production of lower energy neutrinos with a softer low energy spectrum. This would occur either in models with higher compactness (Sikora and Begelman 1992) or indefinitely large nucleon containment times (Szabo and Protheroe 1994). The true situation is probably somewhat intermediate between the various models.

A soft spectrum of low energy neutrinos can also arise from the decay of pions produced in *pp* interactions. Using the X–ray data, we have argued that such interactions should not account for the bulk of the neutrino production (see section 3). In our initial work (Stecker, *et al.* 1991), we estimated that, at most, 10% of AGN neutrinos would be from $pp$ interactions. This conclusion is consistent with the suggested accretion disk production estimate of Nellen, *et al.* (1993).

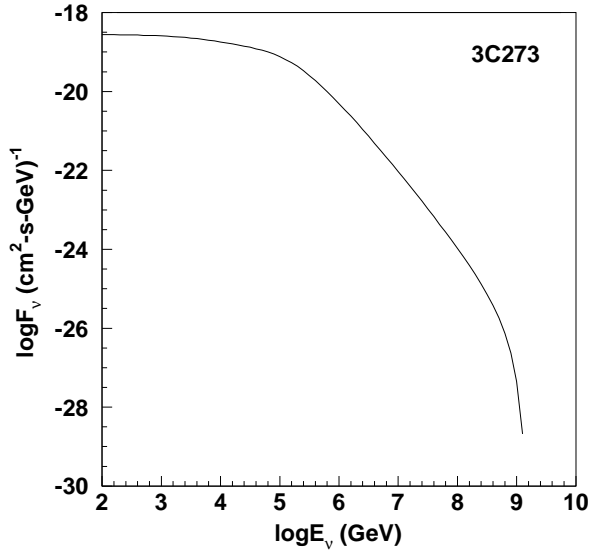

Fig. 1. The predicted $\nu_\mu + \overline{\nu}_\mu$ flux from 3C273, with $L_{\rm x} = 10^{47}$ ergs s$^{-1}$ (Piccinotti, *et al.* 1982) and redshift $z = 0.158$. (A Hubble constant of 50 km-s$^{-1}$-Mpc$^{-1}$ has been assumed in our calculations.) Note that the $\nu_e + \overline{\nu}_e$ flux is half that of the $\nu_\mu + \overline{\nu}_\mu$ flux.

## 6. The Extragalactic Neutrino Background Spectrum

The diffuse $\nu$ flux can be found by integrating over the luminosity function (LF) for quasars. The most recent determinations of the X–ray LF, as well as high redshift luminosity evolution, come from the ROSAT satellite (Boyle, *et al.* 1993). We will adopt model I of Boyle, *et al.* (1993). For this model, the LF at $z = 0$ is given by

$$\rho_o(L_{44})\Delta L_{44} = \begin{cases} 7.6 \times 10^{-7} {\rm Mpc}^{-3} L_{44}^{-1.7}, & L_{\min} < L < L_1 \\ 3.2 \times 10^{-7} {\rm Mpc}^{-3} L_{44}^{-3.4}, & L_1 < L < L_{\max} \end{cases} \quad (3)$$

where the luminosity function $\rho$ is per luminosity interval in units of $(10^{44}$ erg s$^{-1})^{-1}$, $L_{44}$ is the 0.3 to 3.5 keV X-ray luminosity in units of $10^{44}$ erg

s$^{-1}$, $L_{\min} = 10^{39}$ erg s$^{-1}$, $L_1 = 10^{43.8}$ ergs s$^{-1}$ and $L_{\max} = 10^{47}$ ergs s$^{-1}$. These luminosities refer to the X–ray flux in the ROSAT range and must be multiplied by a factor 17 to transform them into a total X–ray flux. The source luminosity as a function of redshift, $z$, is taken to be (Boyle, *et al.* 1993)

$$L(z) = \begin{cases} L(0)(1+z)^{2.73}, & z \leq 2 \\ L(0)3^{2.73}, & z > 2 \end{cases} \quad (4)$$

We calculate the diffuse neutrino background assuming an epoch of formation of quasar activity corresponding to $z_{max} = 4$ and note that our results are not very sensitive to $z_{max}$ in the range between 2 and 4.

Figure 2 shows our new background result, obtained assuming that ~30 % of the quasar X–ray flux is from the $p\gamma$ interactions and subsequent electromagnetic cascading (see section 5). Our results indicate that the diffuse background neutrino flux from quasars could dominate over the atmospheric background in the range $10^5$ to $10^{10}$ GeV.

### 7. Possible Origin of the Broad Line Emission Region

The broad line region (BLR), a characteristic feature of AGN, is thought to be made up of a large number of dense, illuminated clouds (Davidson and Netzer 1979). The large line widths of ~ 5000 km s$^{-1}$ can be explained by doppler shifts due to orbital motions. If the clouds have predominantly circular orbits, then the velocity can be calculated as $v = \sqrt{GM/r}$, or $\beta = \sqrt{R_s/2r}$. For these high velocities, the clouds have to be within ~ $1700R_s$, or $5 \times 10^{16}$ cm for a $10^8$ solar mass AGN black hole.

An outstanding problem in this scenario is the formation and stability of these clouds. While the two phase instability, in which dense condensations can exist in pressure equilibrium with a hot, rarified gas is seen as the best and currently accepted model (Krolik, McKee and Tarter 1981), there are many problems with this explanation (Mathews 1986; Elizur and Ferland 1986).

However, we have previously showed that the high energy neutrino luminosities predicted from quasars will have a profound effect on stars close to the center of the host galaxy, producing stellar winds, swelling the atmospheres and even causing their total disruption (Stecker, *et al.* 1985; Gaisser, *et al.* 1986; Stecker, *et al.* 1991). The column density for absorption of a ~ $10^6$ GeV neutrino is $X = m_H/\sigma_{\nu N} \sim 2 \times 10^9$ g cm$^{-2}$, while total column density for a solar mass star $<\rho> R \sim 10^{12}$ g cm$^{-2}$. Using a conservative disruption criterion that the neutrino energy deposited in the star be greater than its nuclear energy generation, we find a sphere of stellar disruption, $R_{\rm SSD}$, within a radius of

$$R_{\rm SSD} \approx 30 L_{45}^{1/2} \Big(\frac{M_*}{M_O}\Big)^{-1.1} \quad \text{light days} \qquad (5)$$

where $M_*$ and $M_O$ are the masses of the irradiated star and the Sun respectively. This radius is the same as that inferred above for the BLR, and matches well with the current best observational determinations (Netzer 1989; Clavel, *et al.* 1991) and previous calculations have shown that outflowing material from stellar disruption provides, at least qualitatively, an environment similar to that in which the broad lines must arise Kazanas 1989). Thus, high energy neutrinos from an AGN core could play a key role in producing the BLR.

## 8. Neutrinos from the Cores and Jets of Blazars

The evidence for relativistic bulk motions and beaming in blazars ("optically violent variable" quasars and BL Lac objects) implies that the intrinsic luminosities and sizes of the emission regions are more uncertain than in radio quiet objects. However, if the above discussion of the origin of the BLR is correct, then the cores of these objects should similarly produce neutrino fluxes. These objects would be expected to contain the same sort of nucleus as the radio quiet objects, but with the addition of jet emission from further out.

With the observation of $\gamma$-ray emission above 100 MeV from at least 40 blazars out to redshifts greater than 2 by the EGRET group (Fichtel 1994), it has now become possible to estimate the corresponding high energy neutrino emission from the jets of these sources. Let us assume that the observed $\gamma$-rays are produced in the relativistic jets of these objects and are beamed because (a) as discussed before, we do not expect $\gamma$-rays above a few MeV to escape from the core of a typical quasar, and (b) EGRET observed time variability on the order of a day from 3C279, implying relativistic beaming. Let us further assume the the observed $\gamma$-rays are produced by high energy protons, leading to pion production and the subsequent decay of neutral pions into $\gamma$-rays and accompanying charged pions into electrons, positrons and neutrinos. We further note that the observed differential spectrum of 3C279 is consistent with a roughly $E^{-2}$ power law up to energies in excess of 5 GeV (Hartman, *et al.* 1992), as would be expected for shock accelerated proton primaries leading to pion decay $\gamma$-rays. Also, the detection of $\gamma$-rays above 0.5 TeV from Mrk 421 by the Whipple Observatory group (Punch, *et al.* 1992), in combination with the EGRET detection, implies a roughly $E^{-2}$ spectrum for this source up to TeV energies and possibly beyond. Such high energies are hard to account for from electron acceleration alone, since radiative energy losses of electrons at such energies are quite severe.

Mannhiem, *et al.* (1992) have proposed a model for proton acceleration in blazar jets.

If we make the above assumptions, we expect that the ratio of $\gamma$-rays to $\nu_\mu + \overline{\nu}_\mu$ should be 1:1. Assuming an $E^{-2}$ differential spectrum, the ratio above a given energy should be 2:1, since the mean energy of a $\gamma$-ray from pion decay is roughly twice the energy of a neutrino from the decay of a pion of equal energy. We can then predict TeV neutrino fluxes from the EGRET sources. For Mrk 421, with a reported $\gamma$-ray flux above 0.5 TeV of $\sim 1.2 \times 10^{-11}$ cm$^{-2}$ s$^{-1}$, we obtain an immediate predicted neutrino flux above 1 TeV of $\sim 6 \times 10^{-12}$ in these same units. This source is at a low enough redshift (0.031) so that no significant $\gamma$-ray absorption at TeV energies is expected, however, the other EGRET sources are at higher redshifts where pair production interactions with intergalactic infrared photons can cut off the observed $\gamma$-ray spectrum (Stecker, De Jager and Salamon 1992). Indeed, although Mrk 421 is the weakest of the EGRET sources yet reported, none of the other sources searched for were detected by the Whipple team above 0.5 TeV (Kerrick, *et al.* 1993). However, correspondingly high energy neutrinos can reach us unattenuated, so that we may use an extrapolation of the EGRET spectrum to predict TeV neutrino fluxes.

For the jet in 3C279, in its more quiescent state, corresponding to the second EGRET observing period, extrapolation of the $E^{-2}$ $\gamma$-ray spectrum leads to a predicted jet muon neutrino flux of $\sim 4 \times 10^{-11}$ above 1 TeV. For the flaring phase the corresponding neutrino flux would be $\sim 2 \times 10^{-10}$, twice the minimum detectable flux for DUMAND II. Our predicted neutrino flux above 1 TeV for the core of this source would be $\sim 10^{-13}$.

For the 3C273 jet, extrapolating the EGRET spectrum (C. von Montigny, *et al.* 1993) leads to a predicted $\nu_\mu$ flux above 1 TeV of $\sim 10^{-13}$. The corresponding prediction for the core flux is $\sim 4 \times 10^{-14}$.

It is interesting to note that for superluminal jet sources, with a typical bulk Lorentz factor of $\sim 10$, the beamed flux would appear to be $\sim 10^3$ times brighter than that of an isotropic source with the same intrinsic production rate.

Using the 1:1 correspondence of $\gamma$-rays to $\nu_\mu$'s (see above), combined with an extrapolated estimate of the blazar $\gamma$-ray background which we published recently (Salamon and Stecker 1994), we immediately obtain the estimate of the neutrino background flux from blazar jets shown in Figure 2, assuming the spectrum continues as a power law well above TeV energies. Figure 2 also shows the AGN neutrino background fluxes calculated by us here and by other workers.

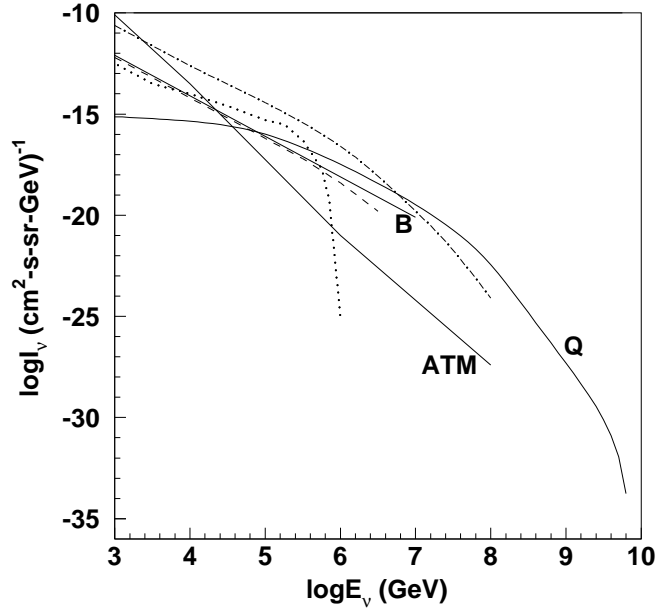

Fig. 2. Our revised integrated high energy $\nu_\mu + \overline{\nu}_\mu$ neutrino background from quasars (Q) and blazars (B) (thick solid lines), as well as the fluxes calculated by Sikora and Begelman (1992) (dotted line), Biermann (1992) (dashed line), and the geometric mean for the $b = 1$ model of Szabo and Protheroe (1994) (dash-dot line). Also shown is the horizontal $\nu_\mu + \overline{\nu}_\mu$ flux from high energy cosmic rays interacting with the Earth's atmosphere (ATM, solid line) (Stecker 1979).

## 9. Discussion

As can be seen from Figure 2, the high energy neutrino background fluxes from AGN calculated by various workers are comparable. This is not too surprising, since they are mainly based on the assumptions of shock acceleration of protons to high energies followed by photomeson production and subsequent charged pion decay. All results are normalized to AGN X-ray or X-ray background energy fluxes in some way.

The results shown in Figure 2 lead to predictions of observable high energy neutrino background fluxes. Indeed, presently existing data are beginning to test the different predictions of the models (Zas, Halzen and Vazquez 1993; Rhode 1994; Aglietta, *et al.* 1994; Gaisser, Halzen and Stanev 1995). In particular, the analyses of upper limits on muon fluxes from the Frejus experiment (Rhode 1994; Gaisser, *et al.* 1995) are consistent with our quasar background flux prediction, but appear to rule out part of the range of model parameters considered by Szabo and Protheroe (1994). Our prediction is also consistent with bounds on giant horizontal air showers established with the AKENO array (Zas, Halzen and Vazquez 1993) and the EAS-TOP collaboration (Aglietta, *et al.* 1994). However, the predicted flux is not very far below the presently established bounds.

The event rates for various $\nu$ detectors can be calculated from our background spectrum given in Figure 2. The Fly's Eye detector, with its energy threshhold of $10^8$ GeV would expect to see $\ll 1$ event per year, consistent with their present non–detection. Their next generation instrument, the HiRes Eye would see $< 1$ *downward* events per year. Above the HiRes energy threshold of $\sim 10^8$ GeV, *upward* neutrinos are severely attenuated by absorption in the Earth. However, the upward event rates at energies greater than $10^5$ GeV for DUMAND II, AMANDA and NESTOR are of order $10^2$ yr$^{-1}$.

We note that $n\gamma$ interactions produce a significant flux of $\overline{\nu}_e$'s at 6.3 PeV, the Glashow resonance energy for the $\overline{\nu}_e e^- \to W^-$ process. An observation of an enhanced event rate at this energy would indicate that AGN are optically thick to neutrons. We estimate that DUMAND II would see on the order of 10 events per year from $W^-$ resonance production. (The reader is referred to the articles in Stenger, *et al.* (1992) for a discussion of various neutrino detector experiments.)

We conclude that results from high energy neutrino detectors either planned or presently under construction should be able to test the viability of the idea that significant proton acceleration in AGN cores ultimately produces a substantial fraction of their observed X-ray luminosity.